\documentclass[iop]{emulateapj}

\shorttitle{Blandford-Znajek revisited} \shortauthors{Contopoulos
et al.}

\def\gsim{\mathrel{\raise.5ex\hbox{$>$}\mkern-14mu
             \lower0.6ex\hbox{$\sim$}}}
\def\lsim{\mathrel{\raise.3ex\hbox{$<$}\mkern-14mu
             \lower0.6ex\hbox{$\sim$}}}

\begin{document}

\author{Ioannis Contopoulos}
\affil{Research Center for Astronomy, Academy of Athens, Athens
11527, Greece; icontop@academyofathens.gr}
\author{Demosthenes Kazanas}
\affil{NASA/GSFC, Code 663, Greenbelt, MD 20771, USA;
demos.kazanas@nasa.gov}
\author{Demetrios B. Papadopoulos}
\affil{Department of Physics, Aristotle University of Thessaloniki,
Thessaloniki 54124, Greece; papadop@astro.auth.gr}
\title{The Force-Free Magnetosphere of a Rotating Black Hole}



\begin{abstract}We revisit the Blandford \& Znajek~(1977) process
and solve the fundamental equation that governs the structure of
the steady-state force-free magnetosphere around a Kerr black
hole. The solution depends on the distributions of the magnetic
field angular velocity $\omega$ and the poloidal electric current
$I$. These are not arbitrary. They are determined
self-consistently by requiring that magnetic field lines cross
smoothly the two singular surfaces of the problem, the inner
`light surface' located inside the ergosphere, and the outer
`light surface' which is the generalization of the pulsar light
cylinder. We find the solution for the simplest possible magnetic
field configuration, the split monopole, through a numerical
iterative relaxation method analogous to the one that yields the
structure of the steady-state axisymmetric force-free pulsar
magnetosphere (Contopoulos, Kazanas \& Fendt 1999). We obtain the
rate of electromagnetic extraction of energy and confirm the
results of Blandford and Znajek and of previous time dependent
simulations. Furthermore, we discuss the physical applicability of
magnetic field configurations that do not cross both `light
surfaces'.
\end{abstract}



\section{Introduction}

The electromagnetic extraction of energy from a spinning black
hole can be one of the most powerful and efficient engines in the
Universe. It was Blandford \& Znajek~(1977) (hereafter BZ77) who
first argued that a spinning black hole threaded by a sufficiently
strong magnetic field and permeated by an electron-positron plasma
generated from pair cascades, establishes a force-free
magnetosphere, in direct analogy to the theory of the axisymmetric
pulsar magnetosphere developed a few years earlier by Goldreich \&
Julian~(1969). They derived the fundamental equation that governs
the structure of this magnetosphere, the general relativistic
force-free Grad-Shafranov equation, and obtained a second order
perturbative solution for small values of the Kerr parameter $a$
and a scaling for the electromagnetic power extracted from the
rotating black hole.

The structure, singular surfaces and general properties of this
equation for magnetic fields threading Schwarzschild and Kerr
black holes were discussed recently by Uzdensky~(2004, 2005). He
emphasized the need for a self-consistent determination of the
distributions of the magnetic field angular velocity and the
poloidal electric current, something that early relaxation methods
(Macdonald 1984) and later higher order perturbative solutions
(Tanabe \& Nagataki 2008) did not take into account. He also noted
that this may be a very difficult task, and indeed, to the best of
our knowledge a solution of this equation in the general case of
an open magnetic field configuration threading a Kerr black hole
is still lacking. Nevertheless, the broader astrophysical interest
in the Blandford-Znajek process, has prompted in the past decade
several researchers to seek and obtain the steady-state structure
of the spinning black hole magnetospheres through time-dependent
general relativistic force-free numerical simulations (e.g.
Komissarov \& McKinney~2007, Palenzuela {\rm et al.}~2011,
Lyutikov \& McKinney~2011). These works established the universal
applicability of the electromagnetic extraction of the rotation
energy of a Kerr black hole in powering active galactic nuclei,
jets in X-ray binaries, and even gamma-ray bursts.

The presence of this extra source of energy in addition to that of
the accretion disk may account for some of the phenomenology. One
should bear in mind, however, that the situation is more
complicated. For instance: (1) Almost $90\%$ of extragalactic
sources are radio quiet, i.e. they do not produce electromagnetic
outflows in the form of radio jets, and yet they are believed to
contain both constituents of the Blandford-Znajek process, namely
a spinning black hole and a strong magnetic field threading it
(e.g. Kukula~2003); (2) The formation and disruption of radio jets
in black hole X-ray binaries over short time scales is at odds
with this general notion (e.g. Belloni~2010); (3) There seems to
exist strong observational evidence that the power in the radio
jet is not directly related to black hole spin, contrary to the
main result of BZ77 (Fender, Gallo \& Russell~2010; however, see
also Narayan \& McClintock~2012). Notice, though, that this may be
due to an erroneous determination of the black hole spin.

The above justify an independent re-evaluation of the
Blanford-Znajek process through the solution of the general
relativistic force-free Grad-Shafranov equation. The special
relativistic form of that equation, the so called pulsar equation,
was only solved in 1999 by Contopoulos, Kazanas and Fendt who
showed that the magnetic field structure can only be obtained
together with the determination of the unique electric current
distribution that allows for a smooth and continuous structure to
exist. In that respect, steady-state solutions offer a deeper
insight in the physics of the problem than numerical
time-dependent ones. It should be noted that, while BZ77 provided
an estimate of the energy extraction efficiency of this process,
they did not obtain an exact solution of the structure of the
black hole magnetosphere. Also, at that time, {\em the critical
role of the singular surfaces and the distributions of the
magnetospheric electric current and field line angular velocity
had not been appreciated}. Today, we revisit this problem with all
the knowledge we carry from our 13-year long investigation of the
force-free pulsar magnetosphere.

\section{The general relativistic pulsar equation}

In order to derive the fundamental equation that governs the
steady-state structure of the force-free magnetosphere around a
Kerr black hole we follow closely the 3+1 formulation of Thorne \&
Macdonald~(1982) used by most researchers in the astrophysical
community (e.g. Uzdensky~2005). We restrict our analysis to
steady-state and axisymmetric space times where
$(\ldots)_{,t}=(\ldots)_{,\phi}=0$. In that case, the general
4-dimensional space-time geometry may be written in
Boyer-Lindquist spherical coordinates $x^\mu\equiv
(t,r,\theta,\phi)$ as
\begin{eqnarray}
{\rm d}s^2 & = & g_{\mu\nu}{\rm d}x^\mu {\rm d}x^\nu\nonumber\\
& = & -\alpha^2{\rm d}t^2+\frac{A\sin^2\theta}{\Sigma}({\rm
d}\phi-\Omega{\rm d}t)^2\nonumber\\
 & & +\frac{\Sigma}{\Delta}{\rm
d}r^2+\Sigma{\rm d}\theta^2\ .\label{st2}
\end{eqnarray}
Here, $\alpha\equiv (\Delta\Sigma/A)^{1/2}$ and $\Omega\equiv
2aMr/A$ are the lapse function and angular velocity of {\it
`zero-angular momentum'} fiducial observers (ZAMOs) respectively,
\[
\Delta\equiv r^2-2Mr+a^2\ \ ,\ \ \Sigma\equiv r^2+a^2\cos^2\theta\
\ ,\
\]
\[A\equiv (r^2+a^2)^2-a^2\Delta\sin^2\theta\ ,
\]
$M$ is the black hole mass, and $a$ its angular momentum
($0\le a\le M$). Throughout this paper we adopt geometric
units where $G=c=1$. Semicolon stands for covariant derivative,
comma for partial derivative. Latin indices denote spatial
components $(1-3)$, Greek indices denote space-time components
$(0-3)$, and `$\sim $' denotes the spatial part of vectors. ZAMOs
move with 4-velocity $U^\mu=(1/\alpha\ ,0\ ,0\ ,\Omega/\alpha)$
orthogonal to hypersurfaces of constant $t$.
The force-free magnetosphere of a spinning black hole is
characterized by the electromagnetic energy-momentum tensor
\begin{equation}\label{e4}
T^{\mu\nu}=\frac{1}{4\pi}(F_{\alpha}^{\mu}
F^{\nu\alpha}-\frac{1}{4}F_{\alpha\beta} F^{\alpha\beta}
g^{\mu\nu})\ ,
\end{equation}
and the condition $T^{\mu\nu}_{;\nu}=0$. Here, the rest mass and
pressure contribution have been neglected. The electromagnetic
field tensor $F^{\mu \nu}$ is related to the electric and magnetic
fields $E^\mu, B^\mu$ measured by ZAMOs through $F^{\mu \nu}=U^\mu
E^\nu -U^\nu E^\mu + \epsilon^{\mu\nu\lambda\rho}B_\lambda U_\rho$
($\epsilon_{\mu\nu\lambda\rho}\equiv
[\mu\nu\lambda\rho]|\mbox{det}(g_{\mu\nu})|^{-1/2}$ is the
4-dimensional Levi-Civita tensor). Under these conditions, the
fundamental equation that governs the steady-state structure of
the force-free magnetosphere around a Kerr black hole becomes
\begin{equation}
\rho_e \tilde{E} + \tilde{J}\times \tilde{B}=0\ . \label{JcrossB}
\end{equation}
$\rho_e$ and $\tilde{J}$ are the electric charge and current
densities respectively. Eq.~(\ref{JcrossB}) is supplemented by
Maxwell's equations of electrodynamics
\begin{eqnarray}\label{k21}
\tilde{\nabla}\cdot \tilde{B} & = & 0\nonumber\\
\tilde{\nabla}\cdot \tilde{E} & = & 4\pi\rho_e\nonumber\\
\tilde{\nabla}\times (\alpha \tilde{B}) & = &
4\pi\alpha\tilde{J}\nonumber\\
\nabla\times (\alpha \tilde{E}) & = & 0 \ .
\end{eqnarray}
Here,
\[
\tilde{\nabla}\cdot \tilde{A} \equiv A^{j}_{;j}\ ,\ \
\tilde{A}\cdot\tilde{B} \equiv g^{ij}A_i B_j\ ,
\]
\[
(\tilde{\nabla}\times \tilde{A})^i \equiv [ijk] |{\rm
det}(g_{lm})|^{-1/2} A_{k;j}\ ,
\]
\[
\ \ (\tilde{A}\times \tilde{B})^i  \equiv [ijk] |{\rm
det}(g_{lm})|^{-1/2}A_{j}B_{k}\ .
\]
For several applications in astrophysics, perfect (infinite)
conductivity is a valid approximation. In this case,
\begin{equation}\label{ideal}
\tilde{E}\cdot \tilde{B}=0\ ,
\end{equation}
and the electric and magnetic vector fields can be expressed in
terms of three scalar functions, $\Psi(r,\theta)$, $\omega(\Psi)$,
and $I(\Psi)$ as
\begin{equation}
\tilde{B}(r,\theta)=\frac{1}{\sqrt{A}\sin\theta}\left\{
\Psi_{,\theta},-\sqrt{\Delta}\Psi_{,r},\frac{2I\sqrt{\Sigma}}{\alpha}\right\}
\label{B}
\end{equation}
\begin{equation}
\tilde{E}(r,\theta)=\frac{\Omega-\omega}{\alpha
\sqrt{\Sigma}}\left\{
\sqrt{\Delta}\Psi_{,r},\Psi_{,\theta},0\right\}\ .
\end{equation}
$\omega$ is the angular velocity of the magnetic field lines, $I$
is the poloidal electric current flowing through the circular loop
$r=$const., $\theta=$const., and $\Psi$ is equal to $(2\pi)^{-1}$
times the total magnetic flux enclosed in that loop. Notice that
the electric field changes sign close to the horizon with respect
to its sign at large distances. As explained in BZ77, a rotating
observer (ZAMO) will in general see a Poynting flux of energy {\it
entering} the horizon, but he will also see a sufficiently strong
flux of angular momentum {\it leaving} the horizon. That ensures
that energy is extracted from the black hole. The poloidal
component of Eq.~(\ref{JcrossB}) then yields the general
relativistic force-free Grad-Shafranov equation
\[
\left\{\Psi_{,rr}+\frac{1}{\Delta}\Psi_{,\theta\theta}
+\Psi_{,r}\left(\frac{A_{,r}}{A}-\frac{\Sigma_{,r}}{\Sigma}\right)
-\frac{\Psi_{,\theta}}{\Delta}\frac{\cos\theta}{\sin\theta}
\right\}
\]
\[\cdot\left[1-\frac{\omega^2
A\sin^2\theta}{\Sigma}+\frac{4M\alpha\omega r\sin^2\theta}{\Sigma}
-\frac{2Mr}{\Sigma}\right]
\]
\[
-\left(\frac{A_{,r}}{A}-\frac{\Sigma_{,r}}{\Sigma}\right)\Psi_{,r}
-\left( 2\frac{\cos\theta}{\sin\theta}-\frac{A_{,\theta}}{A}
+\frac{\Sigma_{,\theta}}{\Sigma}\right) \] \[\cdot(\omega^2
A\sin^2\theta-4M\alpha\omega r\sin^2\theta +2Mr
)\frac{\Psi_{,\theta}}{\Delta\Sigma}
\]
\[
+\frac{2Mr}{\Sigma}\left(\frac{A_{,r}}{A}-\frac{1}{r}\right)\Psi_{,r}
+\frac{4\omega M\alpha r\sin^2\theta}{\Sigma}
\]
\[\cdot\left\{
\Psi_{,r}\left(\frac{1}{r}-\frac{A_{,r}}{A}\right)
-\frac{\Psi_{,\theta}}{\Delta}\frac{A_{,\theta}}{A}\right\}
\]
\begin{equation}
-\frac{\omega'\sin^2\theta}{\Sigma}(\omega A-2\alpha Mr)
\left(\Psi_{,r}^2+\frac{1}{\Delta}\Psi_{,\theta}^2\right) \]
\[=
-\frac{4\Sigma}{\Delta}II' \label{pulsareqGR}
\end{equation}
(Eq.~(3.14) of BZ77 re-written in our notation). Henceforth,
primes will denote differentiation with respect to $\Psi$. One
sees directly that if we set $\alpha=0$ and $M=0$ in
eq.~(\ref{pulsareqGR}) we obtain
\[
\left(\Psi_{,rr}+\frac{1}{r^2}\Psi_{,\theta\theta}
+\frac{2\Psi_{,r}}{r}
-\frac{1}{r^2}\frac{\cos\theta}{\sin\theta}\Psi_{,\theta}\right)\cdot
[1-\omega^2 r^2\sin^2\theta]
\]
\[
-\frac{2\Psi_{,r}}{r} -2\omega^2\cos\theta \sin\theta
\Psi_{,\theta} \]
\begin{equation} -\omega\omega'
r^2\sin^2\theta\left(\Psi_{,r}^2+
\frac{1}{r^2}\Psi_{,\theta}^2\right)=-4II'\ , \label{pulsareq}
\end{equation}
which is the well known pulsar equation (Scharlemann \& Wagoner
1973). The zeroing of the term multiplying the second order
derivatives in eq.~(\ref{pulsareqGR}),
\begin{equation}
1-\frac{\omega^2 A\sin^2\theta}{\Sigma}+\frac{4M\alpha\omega
r\sin^2\theta}{\Sigma} -\frac{2Mr}{\Sigma}=0\ ,
\label{singularity}
\end{equation}
yields the singular surfaces of the problem. We will henceforth
call the singular surfaces {\it `light surfaces'}
(LS)\footnote{These are none other than the {\it
`velocity-of-light surfaces'} of Macdonald~(1984) where the speed
of a particle moving purely toroidally with angular velocity
$\omega$ equals $c$.}. When we set $M=\alpha=0$, it yields the
standard pulsar light cylinder $r\sin\theta=c/\omega$. In general,
the shape of the outer LS is only asymptotically cylindrical as
$\theta\rightarrow 0$ (see figure~2 below), and an inner LS
appears inside the ergosphere. We will henceforth use the notation
`light cylinder' (LC) and `outer LS' interchangeably. It is
interesting to note that the outer boundary of the ergosphere
corresponds to the solution of the singularity condition for
$\omega=0$, whereas the inner boundary (the event horizon at
$r=r_{\rm BH}\equiv M+\sqrt{M^2-a^2}$) corresponds to the solution
of the singularity condition for $\omega=\Omega_{\rm BH}\equiv
a/(r_{\rm BH}^2+a^2)$, where $\Omega_{\rm BH}$ is the angular
velocity of the black hole. It is also interesting to note that
the natural `radiation condition' at infinity (energy must flow
outwards along all field lines) requires that
\begin{equation}
0\leq \omega \leq \Omega_{\rm BH} \label{radiation}
\end{equation}
(BZ77), and therefore indeed the inner LS lies inside the
ergosphere.

Both Eqs.~(\ref{pulsareqGR}) and (\ref{pulsareq}) contain the two
functions, $\omega(\Psi)$ and $I(\Psi)$, which must be determined
by the physics of the problem. In the case of an axisymmetric
spinning neutron star, $\omega$ is usually taken to be equal to
the neutron star angular velocity $\Omega_{\rm
NS}$\footnote{Notice that in the presence of particle acceleration
magnetospheric `gaps', this is not 100\% exact (Ruderman \&
Sutherland~1975, Contopoulos~2005). In particular, in old pulsars
near their death line $\omega \ll \Omega_{\rm NS}$.}. In pulsars,
$I(\Psi)$ is self-consistently determined through an iterative
numerical technique that implements a smooth crossing of the
relativistic Alfv\`{e}n surface, which is the usual light cylinder
where $r\sin\theta=c/\omega$ (Contopoulos, Kazanas \& Fendt~1999,
Timokhin~2006). In the case of a spinning black hole, the
situation is qualitatively similar but more complicated. Contrary
to a neutron star, the black hole does not have a solid surface,
and therefore it cannot impose any restriction on $\omega$ other
than the `radiation condition' Eq.~(\ref{radiation}). The only
natural restriction that is imposed by the physical problem is
that magnetic field lines must be smooth and continuous
everywhere. $\omega(\Psi)$ must therefore be determined together
with $I(\Psi)$ through the condition of smooth crossing of both LS
of the problem, the inner one inside the ergosphere, and the outer
one at large distances. In the next section, we will see that this
can be achieved through an iterative numerical technique analogous
to the one employed in the pulsar magnetosphere. As remarked
previously, iterating with respect to two functions simultaneously
is indeed a very difficult task. This is why Uzdensky~(2005) opted
to consider only field lines that connect the black hole horizon
to the surrounding accretion disk where $\omega(\Psi)$ is
determined by the Keplerian angular velocity of the disk.

There exists one more interesting complication.
Eq.~(\ref{radiation}) applies only to open field lines that cross
the event horizon. In the case of a star, field lines that cross
the stellar surface rotate with the angular velocity of the star,
but they do not cross an event horizon (either because no event
horizon forms, or because they simply avoid the horizon), and the
above restriction does not apply. In the present case, we are
interested in studying the fundamentals of the Blandford-Znajek
process, and therefore, we will consider a black hole `as clean as
possible'. Obviously, a magnetized astrophysical black hole is not
isolated. In that respect, we would like here to study an
astrophysical black hole with the minimum number of extra
elements, namely a surrounding thin disk of matter to hold the
magnetic field through the necessary electric currents and
charges. Charges and currents will be produced through pair
production in the black hole magnetosphere. We will ignore
magnetic field lines that do not cross both the inner and outer LS
because we will have no physical way to determine their $\omega$
and the poloidal electric current $I$ contained inside them. We
will also ignore magnetic field lines that cross the equator
outside the black hole event horizon because such field lines need
a source of poloidal electric current on the equator (if they
rotate, they will need to cross at least one LS where the
condition of smooth crossing will in general require a nonzero
electric current to flow along that line).

Therefore, in the present work we will only study black hole
magnetospheres where all magnetic field lines reach the event
horizon, and all magnetic field lines are open. As argued in
Lyutikov~(2012), such configuration is the most natural end stage
of stellar collapse. Notice once again that it is sustained by
electric currents in an equatorial current sheet. The effect of a
surrounding disk of a certain height possibly threaded by the
return magnetic field (as in the scenario predicted by the Cosmic
Battery; Contopoulos \& Kazanas~1998) will be considered in a
future work.

\newpage
\section{The numerical iterative scheme}

We will here describe the technical numerical details that led to
the solution of Eq.~(\ref{pulsareqGR}). Firstly, we change the
radial variable from $r$ to $R(r)\equiv r/(r+M)$. Our numerical
integration extends from $R_{\rm min}\equiv R(r_{\rm BH})$ (the
event horizon) to $R_{\rm max}=1$ (radial infinity). The  $\theta$
coordinate extends from $\theta_{\rm min}=0$ (the axis of
symmetry) to $\theta_{\rm max}=\pi/2$ (the equatorial plane). We
implemented a $256\times 64$ numerical grid uniform in $R$ and
$\theta$. Notice that this grid has a very high resolution in $r$
around the black hole event horizon where the inner LS lies, but
not as high around the outer LS, and in particular at low $\theta$
and $a$ values.

Secondly, we specify boundary conditions on the axis of symmetry,
the horizon, the equatorial plane, and infinity.
$\Psi(\theta=0)=0$.
$\Psi(r_{\rm BH},\theta)$ is determined by the integration over
$\theta$ of the so called Znajek regularization condition
\begin{equation}
I(\Psi)=-\frac{Mr_{\rm BH}\sin\theta}{r_{\rm
BH}^2+a^2\cos^2\theta}(\Omega_{\rm BH}-\omega)\Psi_{,\theta}
\label{Znajek}
\end{equation}
(Znajek~1977). Notice that this condition is updated as the
distribution $I(\Psi)$ is updated along with the numerical
solution for $\Psi(r,\theta)$. The equatorial boundary condition
is interesting. As we discussed in the previous section, the
equatorial region may contain an accretion disk, with or without
its own magnetic field. Before investigating such a complex
astrophysical system, we would like to understand first the
simplest case, that of a magnetized black hole where the electric
currents supporting its magnetic field are distributed on a thin
equatorial current sheet. In that case, the equatorial boundary
condition becomes $\Psi(r,\theta=\pi/2)=\Psi_{\rm max}$. The outer
boundary condition is also not important since the outer boundary
has practically been moved to infinity. In practice, we chose
$\Psi_{,r}(R=1)=0$, but any other boundary condition yields the
same results. We initialize our numerical grid with a split
monopole field configuration
\begin{equation}
\Psi(r,\theta)=\Psi_{\rm SM}(r,\theta)\equiv\Psi_{\rm
max}(1-\cos\theta)\ , \label{splitmonopole1}
\end{equation}
with
\[
\omega(\Psi)=0.5\Omega_{\rm BH}\ \ \mbox{and}
\]
\begin{equation}I(\Psi)=I_{\rm SM}(\Psi)\equiv -0.25\Omega_{\rm
BH}\Psi[2-(\Psi/\Psi_{max})]\ . \label{splitmonopole2}
\end{equation}
The reader can directly check that Eqs.~(\ref{splitmonopole1}) \&
(\ref{splitmonopole2}) are an exact analytical solution of the
pulsar equation (Eq.~\ref{pulsareq}; Michel 1982). Notice the
negative sign in Eq.~(\ref{splitmonopole2}).  For the particular
magnetic field configuration where $B^r$ on the axis is aligned
with the black hole spin direction, this corresponds to an outflow
of electrons. Different initial configurations yield the same
final solution. We discretize all physical quantities on our
$256\times 64$ $(R,\theta)$ grid and we update $\Psi(R,\theta)$
through simultaneous overrelaxation with Chebyshev acceleration
(subroutine SOR from Numerical Recipes; Press, Flannery \&
Teukolsky~1986).

Thirdly, we update the distributions of $\omega(\Psi)$ and
$I(\Psi)$ as follows: At each latitude $\theta$, we check where
the singularity condition (Eq.\ref{singularity}) is satisfied in
$r$. At each such radial position, we extrapolate $\Psi$ inwards
from larger $r$ ($\Psi(r^+,\theta)$) and outwards from smaller $r$
($\Psi(r^-,\theta)$). In general, $\Psi(r^+,\theta)$ and
$\Psi(r^-,\theta)$ differ. Then, at the inner LS we implement
\begin{eqnarray}
\omega_{\rm new}(\Psi_{\rm new}) & = & \omega_{\rm old}(\Psi_{\rm
new})\nonumber \\
& & - 0.1[\Psi(r^+,\theta)- \Psi(r^-,\theta)]\ ,
\end{eqnarray}
whereas at the outer LS we implement
\begin{eqnarray}
I_{\rm new}(\Psi_{\rm new}) & = & I_{\rm old}(\Psi_{\rm new})
\nonumber\\
& & +
0.1[\Psi(r^+,\theta)- \Psi(r^-,\theta)]\nonumber\\
\omega_{\rm new}(\Psi_{\rm new}) & = & \omega_{\rm old}(\Psi_{\rm
new})\nonumber\\
& &- 0.1[\Psi(r^+,\theta)- \Psi(r^-,\theta)]\ ,
\end{eqnarray}
where
\begin{equation}
\Psi_{\rm new} \equiv  0.5[\Psi(r^+,\theta)+ \Psi(r^-,\theta)]
\end{equation}
at each LS. The reasoning here is that we impose weighted
corrections on $\omega(\Psi)$ and $I(\Psi)$ based on the
non-smoothness of the $\Psi(r,\theta)$ distribution along all grid
points inside and outside the two LS, and not through the
regularization conditions as we did in Contopoulos, Kazanas \&
Fendt~(1999). This is a very general procedure that may be applied
to any similar singular equation.

Our numerical scheme does not converge easily, and its parameters
need to be {\it empirically adjusted} by following the progress of
the iteration. The signs of the weighting coefficients are
determined by trial and error: when we choose the wrong signs, the
mismatches $|\Psi(r^+,\theta)- \Psi(r^-,\theta)|$ increase and the
$\Psi(r,\theta)$ distribution is very quickly destroyed. Their
magnitudes are also obtained empirically: too large values (around
unity) lead to over-corrections and instability, too small values
(below~0.01) do not correct fast enough for the iteration to
converge. Furthermore, in order to facilitate convergence, at the
inner LS we update only $\omega$, whereas at the outer LS we
update both $I$ and $\omega$ every other relaxation iteration for
$\Psi$. Also, in order to avoid numerical instabilities, we smooth
out the distributions of $I$ and $\omega$ every 50 relaxation
iterations for $\Psi$. We found that more frequent smoothing
inhibits convergence. Our goal is to minimize the residuals in the
discretized form of eq.~(\ref{pulsareqGR}) together with the
non-smoothness in $\Psi(r,\theta)$, but our method fails for poor
grid resolutions at the inner and/or the outer LS (see Discussion
section).

\begin{figure}
\centering
\begin{minipage}[b]{0.5\linewidth}
\includegraphics[trim=0cm 3cm 0cm 3cm, clip=true,
width=6cm, angle=270]{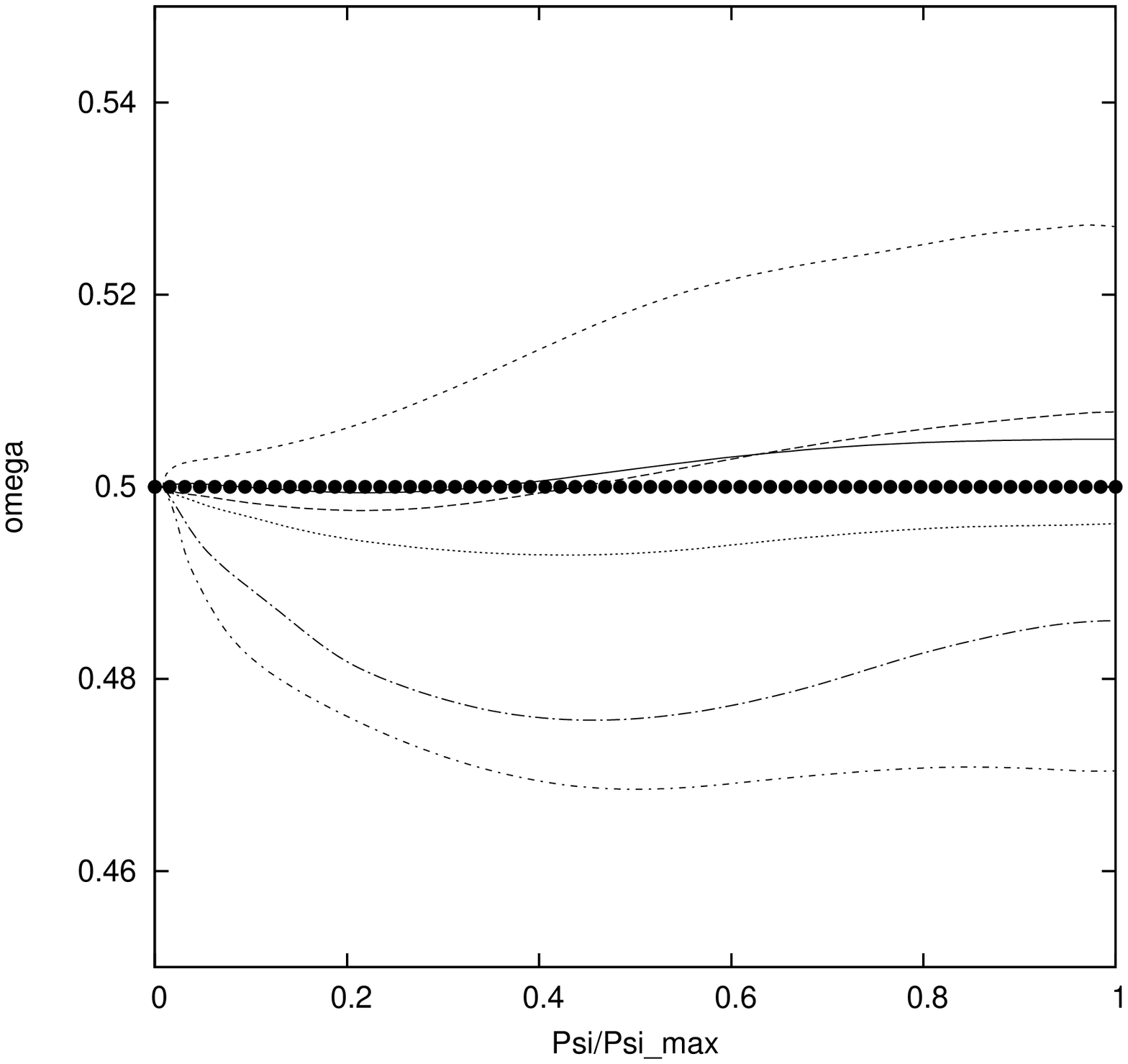}

\centering
\includegraphics[trim=0cm 3cm 0cm 3cm, clip=true,
width=6cm, angle=270]{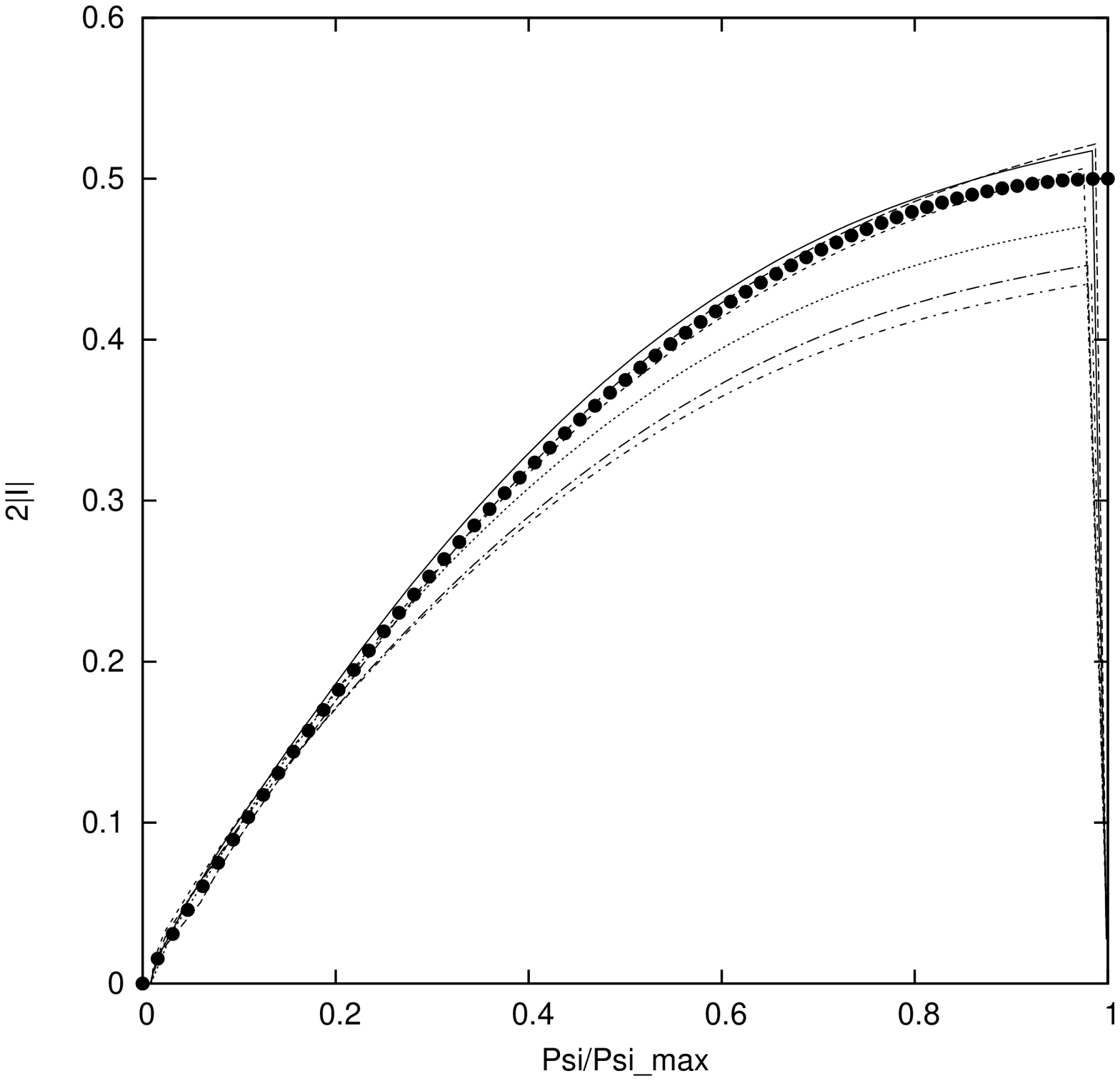}

\centering
\includegraphics[trim=0cm 3cm 0cm 3cm, clip=true,
width=6cm, angle=270]{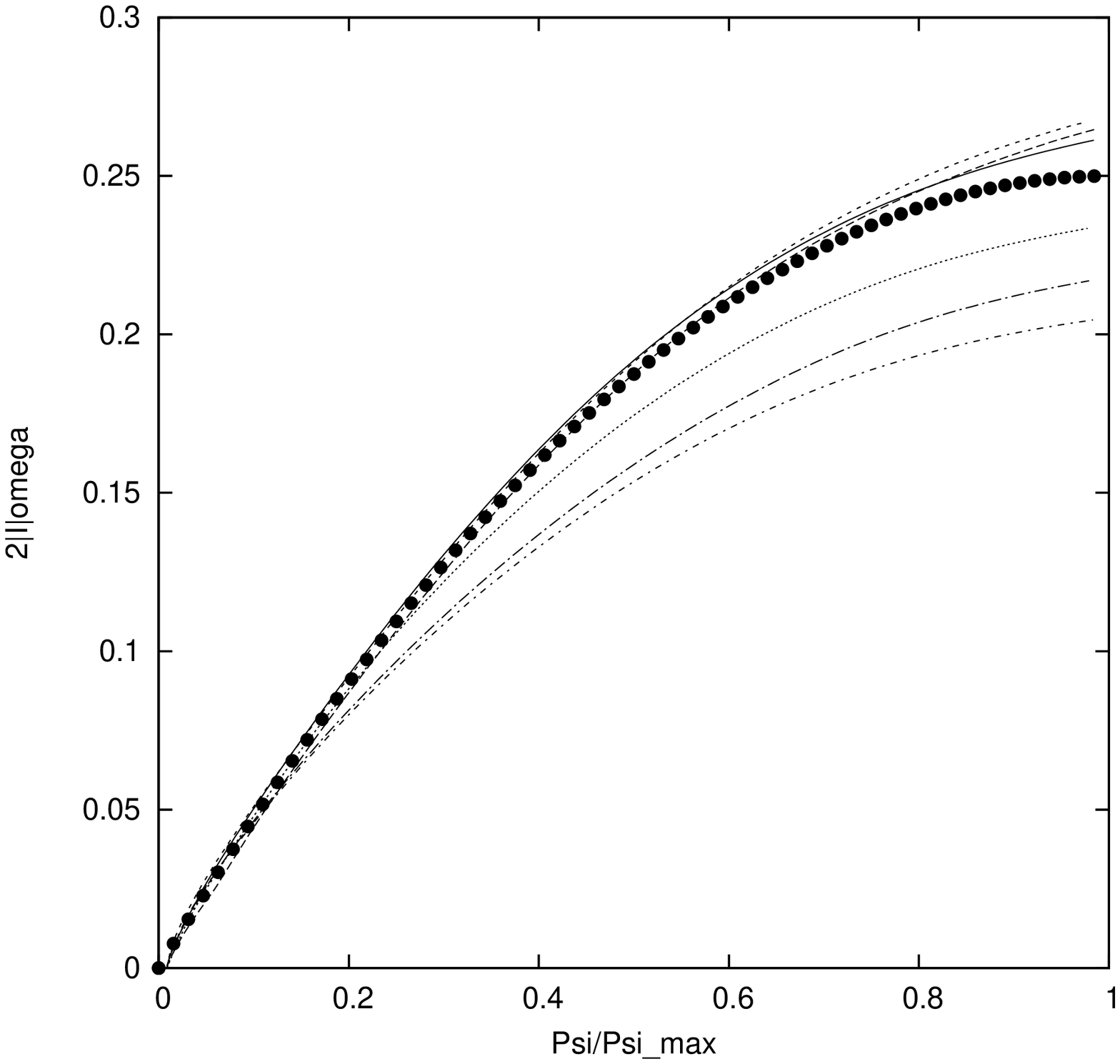}
\end{minipage}

\caption{Distributions of $\omega(\Psi)$ normalized to
$\Omega_{\rm BH}$ (top panel), $2|I(\Psi)|$ normalized to
$\Omega_{\rm BH}\Psi_{max}$ (center panel), and Poynting flux at
large distances $2|I(\Psi)|\omega(\Psi)$ normalized to
$\Omega_{\rm BH}^2\Psi_{max}$ (bottom panel) for various values of
the black hole spin parameter $a/M=0.7$ (solid), 0.8 (short
dashed), 0.9 (dotted), 0.99 (small dots), 0.999 (dash dotted),
0.9999 (short dash dotted). All three distributions are close to
the analytical split monopole expressions of
Eq.~(\ref{splitmonopole2}) (thick dots).}
\end{figure}
\begin{figure}
\begin{minipage}[b]{0.5\linewidth}
\centering
\includegraphics[trim=2cm 4cm 1cm 6cm,
clip=true, width=6cm, angle=270]{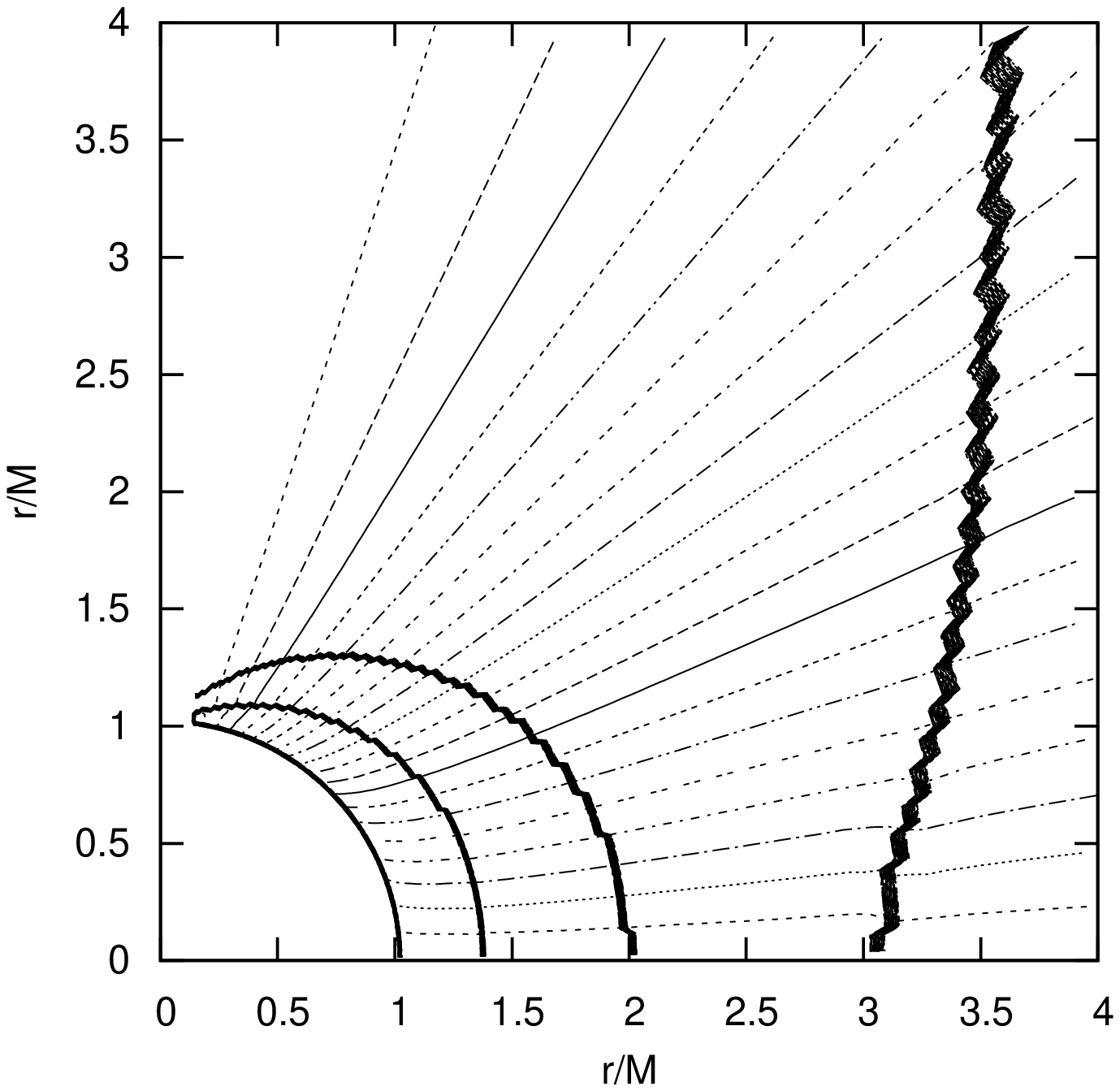}
\end{minipage}
\begin{minipage}[b]{0.5\linewidth}
\centering
\includegraphics[trim=2cm 4cm 1cm 6cm,
clip=true, width=6cm, angle=270]{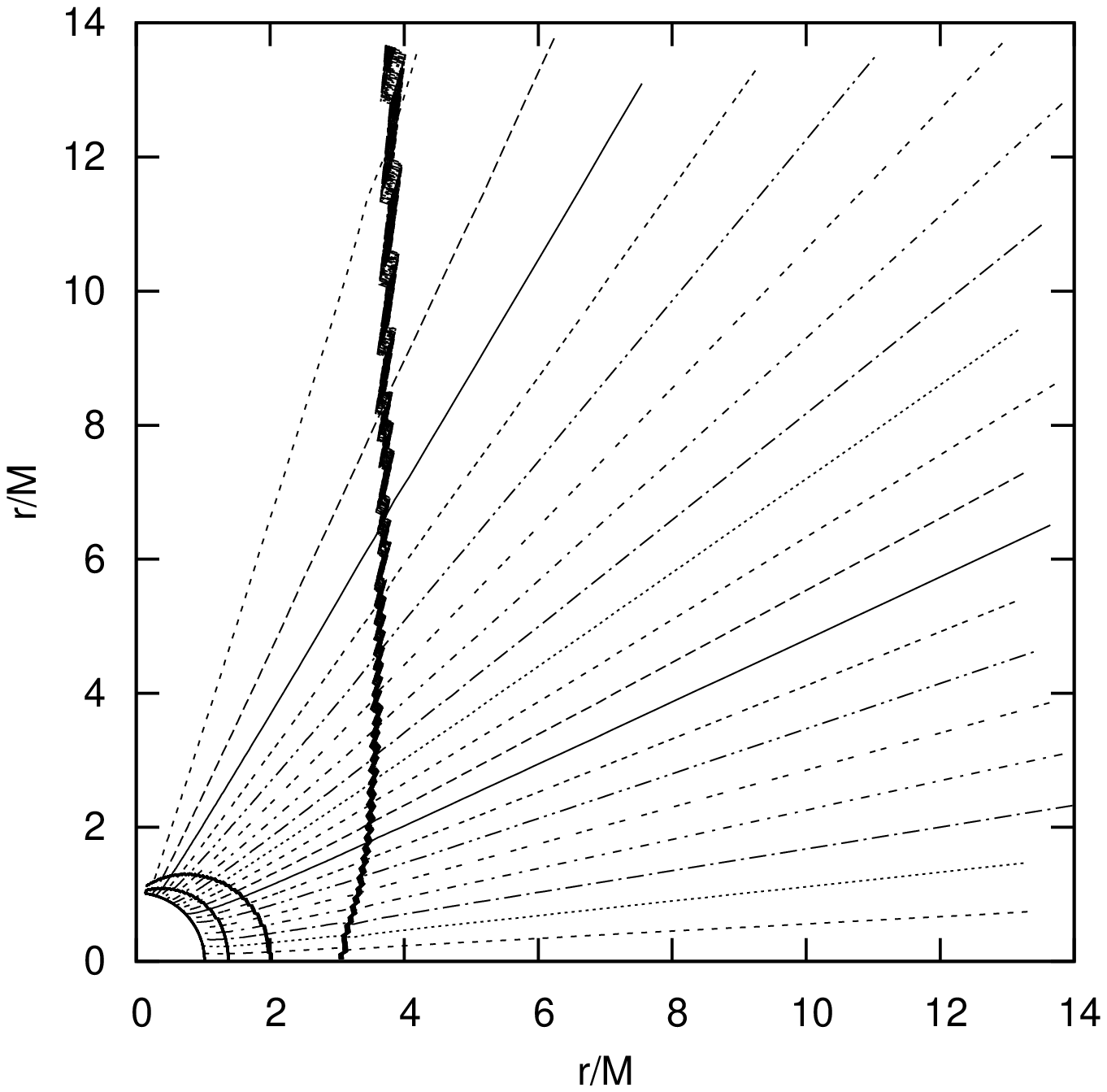}
\end{minipage}
\caption{Poloidal magnetic field lines (lines of constant $\Psi$
in 20 equal subdivisions of $\Psi_{max}$) for $a/M=0.9999$ close
to the event horizon (top panel) and further out (bottom panel).
The thick lines are the event horizon, the inner LS, the outer
boundary of the ergosphere, and the outer LS (from the center
outwards respectively).}
\end{figure}

The results of our numerical integration are shown in figure~1. We
show here the distributions of $\omega(\Psi)$ (top panel),
$2|I(\Psi)|$ (center panel), and Poynting flux at large distances
$2|I(\Psi)|\omega(\Psi)$ (bottom panel) after 6000 relaxation
steps for $\Psi(R,\theta)$, for six values of the black hole spin
parameter $a$. As is the case in pulsars, the magnetospheric
electric current is non-zero at $\Psi=\Psi_{max}$, and the global
electric circuit closes through an equatorial current sheet.
Notice that we have no way to update $\omega$ along the axis, and
therefore we have chosen $\omega(\Psi=0)= 0.5\Omega_{\rm
BH}$.\footnote{We have also run our simulations with
$\omega'(\Psi=0)=0$ and the solutions were similar, only the
convergence of $\omega$ became more unstable.} On the equator we
implemented $\omega'(\Psi=\Psi_{max})=I'(\Psi=\Psi_{max})=0$. The
reader should check the qualitative agreement with the results of
the time-dependent force-free general relativistic simulations
performed by Tchekhovskoy, Narayan \& McKinney~(2010) (their
figure~4).

In figure~2 we show the characteristic magnetic field
configuration for $a/M=0.9999$ near the black hole horizon (left
panel) and at larger distances (right panel). We also plot the
inner LS inside the ergosphere, and the outer LS which becomes
asymptotically cylindrical as $\theta\rightarrow 0$. Notice that
magnetic field lines are monopolar at large distances (beyond a
few times the radius of the event horizon), and they bend upwards
towards the axis as they approach the horizon. This effect is also
seen in the numerical simulations of Lyutikov \& McKinney~(2011)
for $a=0.99$. It is not discernable in the field configurations of
figure~1 of Tchekhovskoy, Narayan \& McKinney~(2010) because the
scale is not adequate, but is implied in their figure~7. Notice
that $\omega(\Psi)$ is equal to $0.5\Omega_{\rm BH}$ to within
10\%, and that $I(\Psi)$ is very close to the split monopole
expression $I_{\rm SM}(\Psi)$ also to within 10\%
(Eqs.~\ref{splitmonopole1} \& \ref{splitmonopole2}). Therefore,
the Znajek condition (Eq.~\ref{Znajek}) can be rewritten using
Eq.~(\ref{B}) as
\[
\left. B^r\right|_{r=r_{\rm BH}} \approx
\frac{\Psi_{max}}{2Mr_{\rm BH}}\frac{\Sigma}{\sqrt{A}}
\]
\begin{equation}=
\frac{\Psi_{max}}{(2M)^2}
\left[1+\frac{(a/M)^2}{[1+\sqrt{1-(a/M)^2}]^2}\cos^2\theta\right]\
.
\end{equation}

Finally, we calculate the total rate of electromagnetic
energy extraction (from both hemispheres) as measured at infinity
through Eq.~(4.11) of Blanford \& Znajek~(1977) as
\begin{eqnarray}
{\cal E} & = &  -2\times 2\pi\int_{\Psi=0}^{\Psi_{max}}
2I(\Psi)\omega(\Psi){\rm d}\Psi \nonumber\\
& = & {\it k} {\cal E}_{\rm SM}\equiv {\it k}
\frac{2\pi}{3}\Omega_{\rm BH}^2\Psi_{max}^2\ .
\end{eqnarray}
We have here normalized our results to the split monopole value
with $\omega_{\rm SM}=0.5\Omega_{\rm BH}$. ${\it k}\approx 0.8$
for $a/M=0.9999$ and is practically indistinguishable from unity
for $a/M\lsim 0.9$ (see bottom panel of figure~1), thus confirming
the main result of BZ77, namely that, in an open magnetic field
configuration with an infinitely thin surrounding equatorial disk,
electromagnetic energy is extracted at a rate proportional to both
$\Omega_{\rm BH}^2$ and $\Psi_{max}^2$.


\section{Discussion}

We believe that the study of the steady-state open force-free
black hole magnetosphere offers a deeper insight in the physics of
the problem than time-dependent numerical simulations. In analogy
to the axisymmetric force-free pulsar magnetosphere, we argued
that steady-state solutions can only be obtained through a
self-consistent determination of the distributions of both the
magnetic field angular velocity and the magnetospheric electric
current $\omega(\Psi)$ and $I(\Psi)$ respectively. {\it These can
only be determined when open field lines cross both singular LS,
the inner LS inside the ergosphere outside the black hole horizon,
and the outer LS which is a deformed light cylinder}. We developed
an iterative numerical procedure analogous to the one we
implemented in the study of the pulsar magnetosphere. Iterating
with respect to two functions simultaneously is a very difficult
task and our method may certainly be improved in the future. Yet,
we are confident enough to suggest that the distributions are
unique for the particular boundary conditions of the problem (all
magnetic field lines thread the black hole horizon, all are open,
and the surrounding disk is infinitely thin) since different
initial conditions for $\Psi$, $\omega$ and $I$ evolve towards the
same final solution (for each value of $a$ that is). Our solutions
may serve as test cases for time-dependent numerical codes. We
plan to apply our method in the study of various astrophysically
interesting situations where the surrounding disk is not
infinitely thin but expands vertically with distance.

All field lines that cross the black hole horizon are bound to
cross the inner LS. This is not the case, though, for the outer
LS. In fact, several numerical simulations exist in the literature
where the black hole is threaded by a vertical uniform magnetic
field, and yet they converge to certain distributions of
$\omega(\Psi)$ and $I(\Psi)$ (e.g. Komissarov  \& McKinney~2007,
Palenzuela {\it et al.}~2011). We expect that in that case, the
solution is mathematically indeterminate, and the solutions shown
depend on the particular choice of outer numerical boundary
conditions. For example, the lateral boundary conditions in
Palenzuela {\em et al.}~(2011) are not perfectly absorbing but
reflect outgoing waves as is manifested in the oscillations seen
in their figure~2. We plan to study such magnetic field
configurations in a future work and check whether we can indeed
freely specify one of the two free functions, $\omega(\Psi)$ or
$I(\Psi)$. We will thus check whether the results of previous
time-dependent numerical simulations are unique or not.

We would like to caution the reader about certain numerical
complications that arise in all numerical simulations of the
spinning black hole magnetosphere, time dependent and steady
state. Numerical simulations that do not adequately resolve both
the inner and outer LS should not be trusted. We faced a similar
problem in our study of the pulsar magnetosphere. In that case,
there exists no inner LS, and in order to study the transition
through the outer LS (the light cylinder) we were forced to
implement our inner boundary condition at a radial distance more
than a hundred times larger than the actual neutron star radius
(Spitkovsky~2006, McKinney~2006, Contopoulos \&
Kalapotharakos~2010). This, however, posed no problem since the
angular velocity of the field lines is set by the rotation of the
pulsar. The black hole problem is more complicated because the
black hole surface is not infinitely conducting, and the numerical
code must adequately resolve both the inner boundary and the light
cylinder. Notice that our numerical method works best as $a/M$
approaches unity since in that case the inner LS is well separated
from the event horizon, and the outer LS is at its closest
approach to the origin of the numerical grid where its radial
resolution is the highest. In that case, a numerical resolution of
256 radial grid points corresponds to 34 grid cells between the
horizon and the inner LS, and 94 grid cells between the two LS
along the equator. When $a/M\gsim 0.9$, our results are rather
resolution independent. On the other hand, as $a/M$ approaches
zero, the ergosphere shrinks in width and the inner LS approaches
so close to the event horizon that our numerical grid does not
have enough radial resolution for the iterative procedure that
yields $\omega(\Psi)$ to properly work\footnote{The solutions for
$a/M=0.7$ \& 0.8 were obtained with a radial grid resolution of
1024. For even smaller values of $a$, our numerical integration
fails because the inner LS is practically indistinguishable from
the event horizon. We can bypass this problem by choosing the
inner boundary condition some distance outside the black hole
horizon, set $\omega\approx 0.5\Omega_{\rm BH}$, and iterate only
with respect to the distribution of poloidal electric current
$I(\Psi)$ that allows for a smooth crossing of the outer LS.}.
At the same time, as $a/M$ decreases so does $\Omega_{\rm BH}$,
pushing the outer LS so far out that our radially expanding
numerical grid has very low radial resolution at the position of
the outer LS for the iterative procedure to properly converge.
Similarly, the resolution of our radially expanding grid on the
light cylinder becomes worse as $\theta\rightarrow 0$, and this
too complicates the convergence of our iterative method. Other
numerical approaches certainly face similar complications at short
and/or at large distances. Our method may be improved by a
reformulation in radially expanding cylindrical coordinates, in
Kerr-Schild coordinates, or in standard spherical coordinates with
a numerical integration outer radial boundary.

In summary, the force-free spinning black hole magnetosphere is
not very different from the force-free axisymmetric pulsar
magnetosphere. In fact, beyond a distance on the order of a few
times the black hole radius, the two problems are almost
identical. An important corollary of this similarity is that {\it
a spinning black hole surrounded by a thin equatorial plasma disk
does not generate a collimated outflow, only an uncollimated wind}
(as in Contopoulos, Kazanas \& Fendt~1999).
Thus, the problem of the black hole jet
formation remains open. A possible resolution must involve a
surrounding thick disk, or a surrounding magnetic disk wind that
gradually collimate the initially monopolar black hole wind into a
collimated jet outflow, as seen in the numerical simulations of
Tchekhovskoy, Narayan \& McKinney~(2010). It is interesting to
note here, again similarly to pulsars, that beyond the light
cylinder {\it the Lorentz factor $\Gamma$ in the monopolar wind
increases linearly with cylindrical distance} as
\begin{equation}
\Gamma \propto \frac{r\sin\theta}{c/\omega} \approx
\frac{r\Omega_{\rm BH}\sin\theta}{2c} \label{Gamma}
\end{equation}
(Contopoulos \& Kazanas~2002), possibly up to the fast
magnetosonic point of the outflow. Therefore, according to
Eq.~(\ref{Gamma}), collimated field/flow lines that do not extend
as far from the axis as uncollimated ones are expected to reach
lower Lorentz factors. Of course, the final Lorentz factor of such
outflows will also depend on the mass loading of a given magnetic
field lines, but such considerations are beyond the scope of the
present work.
We plan to continue our investigation of the
interrelation between the black hole and disk outflows in the
future.

\acknowledgements

We would like to thank Prs. George Contopoulos and Maxim Lyutikov
for interesting discussions. This work was supported by the
General Secretariat for Research and Technology of Greece and the
European Social Fund in the framework of Action `Excellence'.

{}


\begin{thebibliography}{}
\bibitem{B10} Belloni, T. M. 2010, Lec. Notes Phys., {\bf 794}, 53
\bibitem{BZ77} Blandford, R. D. \& Znajek, R. L. 1977, MNRAS, {\bf 179},
433 (BZ77)
\bibitem{C05} Contopoulos, I. 2005, A\& A, {\bf 442}, 579
\bibitem{CK98} Contopoulos, I. \& Kazanas, D. 1998, ApJ, {\bf 508}, 859
\bibitem{CK98} Contopoulos, I. \& Kazanas, D. 2002, ApJ, {\bf 566}, 336
\bibitem{CKF99} Contopoulos, I., Kazanas, D. \& Fendt, C. 1999, ApJ, {\bf 511}, 351
\bibitem{CK10} Contopoulos, I. \& Kalapotharakos, C. 2010, MNRAS, {\bf 404}, 767
\bibitem{FGR10} Fender, R. P., Gallo, E. \& Russell, D. 2010, MNRAS, {\bf 406}, 1425
\bibitem{GJ69} Goldreich, P. \& Julian, W. H. 1969, ApJ, {\bf 157}, 869
\bibitem{KMcK07} Komissarov, S. S. \& McKinney, J. C. 2007, MNRAS, {\bf 377}, L49
\bibitem{K03} Kukula, M. J. 2003, New Astro. Rev., {\bf 47}, 215
\bibitem{L12} Lyutikov, M. 2012, arXiv:1209.3785
\bibitem{LMcK11} Lyutikov, M. \& McKinney, J. C. 2011, Phys. Rev. D., {\bf 84}, 4019
\bibitem{M84} Macdonald, D. A. 1984, MNRAS, {\bf 211}, 313
\bibitem{McK06} McKinney, J. C. 2006, MNRAS, {\bf 368}, L30
\bibitem{M82} Michel, F. C. 1982, Rev. Mod. Phys., {\bf 54}, 1
\bibitem{NMcC12} Narayan, R. \& McClintock, J. E. 2012, MNRAS, {\bf 419}, 69
\bibitem{PBLR11} Palenzuela, C., Bona, C., Lehner, L. \& Reula, O. 2011, Class. Quant. Grav., {\bf 28}, 4007
\bibitem{PFTV86} Press, W. H., Flannery, B. P., Teukolsky, S. A. 1986, `Numerical recipes. The art of scientific computing', Cambridhe Univ. Press
\bibitem{RS75} Ruderman, M. A. \& Sutherland, P. G. 1975, ApJ, {\bf 196}, 51
\bibitem{SW73} Scharlemann, E. T. \& Wagoner, R. V. 1973, ApJ, {\bf 182}, 951
\bibitem{S06} Spitkovsky, A. 2006, ApJ, {\bf 648}, 51
\bibitem{TMcD82} Kip S.Thorne, K. S. \& Macdonald, D. 1982, MNRAS, {\bf 198}, 339
\bibitem{TN08} Tanabe, K. \& Nagataki, S. 2008, Phys. Rev. D, {\bf 78}, 024004
\bibitem{TNMcK10} Tchekhovskoy, A., Narayan, R. \& McKinney, J. C. 2010, ApJ, {\bf 711},
50
\bibitem{T06} Timokhin, A. N. 2006, MNRAS, {\bf 368}, 1055
\bibitem{U05} Uzdensky, D. A. 2004, ApJ, {\bf 603}, 652
\bibitem{U05} Uzdensky, D. A. 2005, ApJ, {\bf 620}, 889
\bibitem{Z77} Znajek, R. L. 1977, MNRAS, {\bf 179}, 457
\end{thebibliography}
\end{document}